\documentclass[authoryear]{elsarticle}
\usepackage{pdfpages}

\usepackage{import, graphicx, xcolor, calc}
\usepackage{subfigure}
\usepackage{hyperref}
\usepackage{float}
\usepackage{svg}
\usepackage{listings}
\usepackage{xcolor}

\hypersetup{
    pdftitle={Volume and Surface Area of two Orthogonal, Partially Intersecting Cylinders: A Generalization of the Steinmetz Solid},
    pdfauthor={Fynn Jerome Aschmoneit, Bastiaan Cockx},
    pdfkeywords={Cylinder intersection, Steinmetz solid, Bicylinder, Geometric integration, Computational Geometry}
}

\usepackage[fleqn]{amsmath}

\journal{-}

\floatstyle{plaintop}
\restylefloat{table}

\usepackage[ left=2.5cm, right=2.5cm, top=2.5cm, bottom=2.5cm, includehead]{geometry}

\begin{document}

\begin{frontmatter}

\title{Volume and Surface Area of two Orthogonal, Partially Intersecting Cylinders: \\ A Generalization of the Steinmetz Solid}

\author[A]{Fynn Jerome Aschmoneit\corref{cor1}}
\author[B]{Bastiaan Cockx}

\affiliation[A]{organization={Department of Mathematical Science, Aalborg University, Copenhagen},
               country={Denmark}}

\affiliation[B]{organization={Bundesamt für Materialforschung und -prüfung}, country={Germany}}

\cortext[cor1]{fynnja@math.aau.dk}
    
\begin{abstract}
    The intersection of two orthogonal cylinders represents a classical problem in computational geometry with direct applications to engineering design, manufacturing, and numerical simulation. 
    While analytical solutions exist for the fully intersecting case—the Steinmetz solid—partial intersections with arbitrary depth ratios $\delta = H/D$ require numerical methods or approximations. 
    This work presents general integral expressions for both the intersection volume and surface area as explicit functions of the intersection depth. 
    Accompanying these exact formulations are empirical approximation functions of the form $V^\prime_{approx}(\delta) = \frac{1}{3}(1 - \cos(\delta\pi))$ and $A^\prime_{approx}(\delta) = 4\sin(\delta\pi/2)$, which provide closed-form evaluations with relative errors below 15\% across the full range $0 \leq \delta \leq 1$. Validation against Quasi-Monte Carlo simulation confirms the accuracy of both the analytical and approximate solutions.
\end{abstract}


\end{frontmatter}

\begin{figure}[H]
    \centering
    \includegraphics[width=0.5\linewidth]{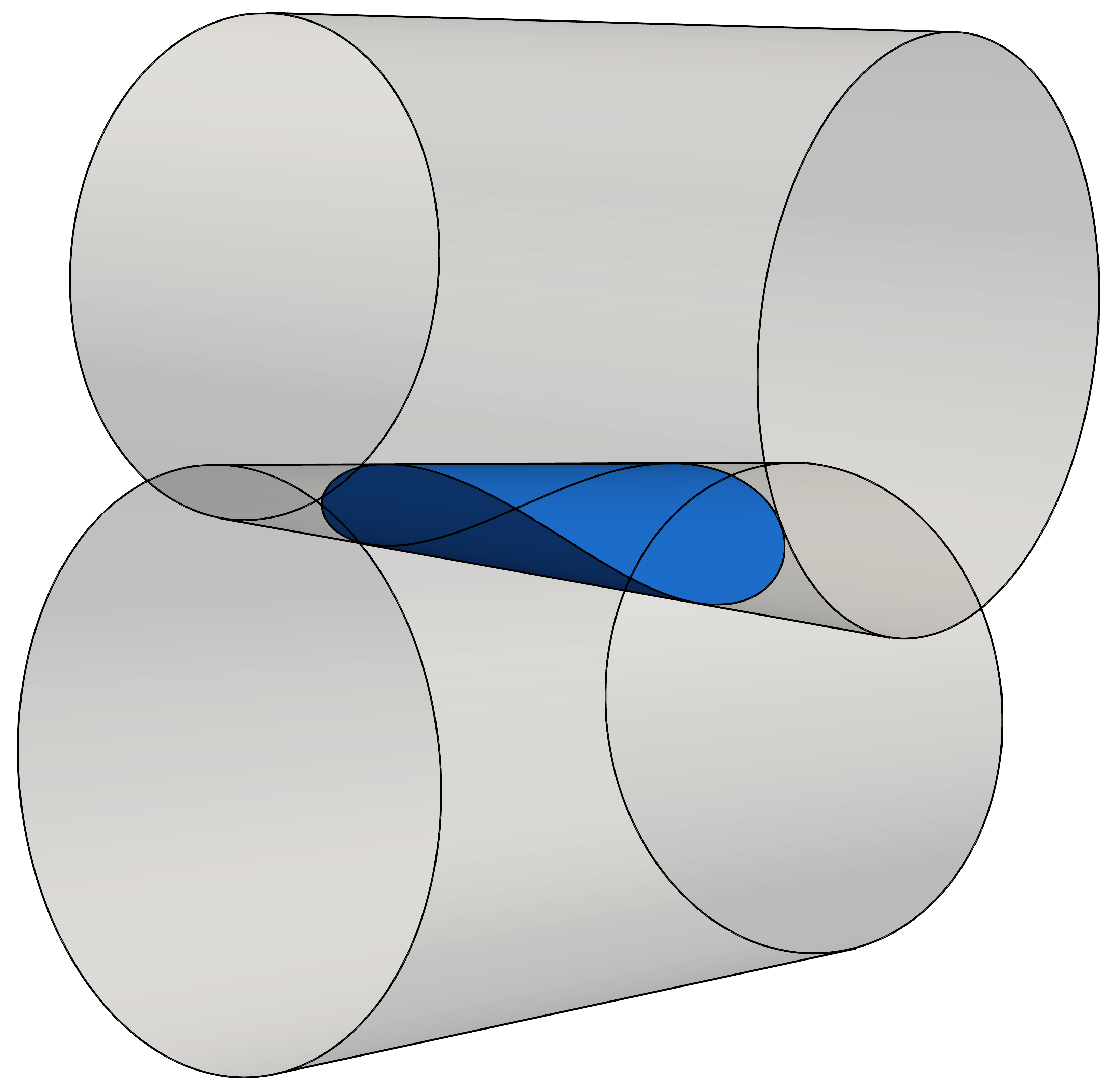}
    \label{fig:intersectingCylinders}
\end{figure}


\section{Introduction}

    The intersection of two cylinders is a classical problem in computational geometry with applications in engineering, manufacturing, and numerical simulation. 
    Determining the volume and surface area of the intersection shape requires careful geometric analysis and integration techniques. 
    Analytical solutions exist only for the fully intersecting case (Steinmetz solid), where both cylinders completely overlap; partial intersections with arbitrary depth ratios necessitate numerical methods or approximation schemes. 
    This work derives general integral expressions for both the intersection volume and surface area as functions of the intersection depth parameter $\delta = H/D$, alongside practical empirical approximation functions that allow for rapid evaluation without numerical integration.
    The analytical results are validated through Quasi-Monte Carlo estimation, establishing the accuracy and applicability of the derived expressions.
    
    \begin{figure}[H]
        \centering
        \includegraphics[width=\linewidth]{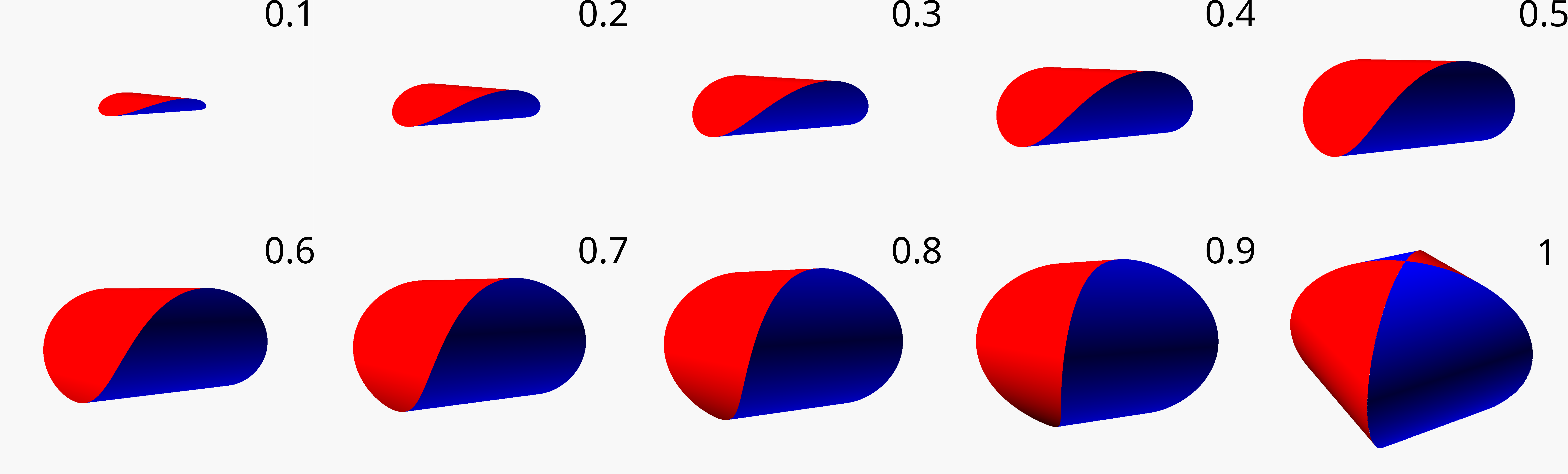}
        \caption{The intersection depth is measured by $\delta = H/D$, which is indicated above each volume. The intersection volume of fully-intersecting cylinders is called a Steinmetz solid ($\delta=1)$.}
        \label{fig:intersectingCylindersSeries}
    \end{figure}

\section{Geometric Integration}
\subsection{Analytical Derivation of Intersection Volume}
    A Cartesian coordinate system is introduced with the $X$ and $Z$ axes aligned parallel to the bottom and top cylinders, respectively.
    The $Y$ axis is oriented in the vertical direction.   
    The origin is placed in the center of the intersection shape.
    The intersection height, i.e., the extent of the intersection in $Y$ direction, is denoted by $H$.
    
    \begin{figure}[H]
        \centering
        \includegraphics[width=0.4\linewidth]{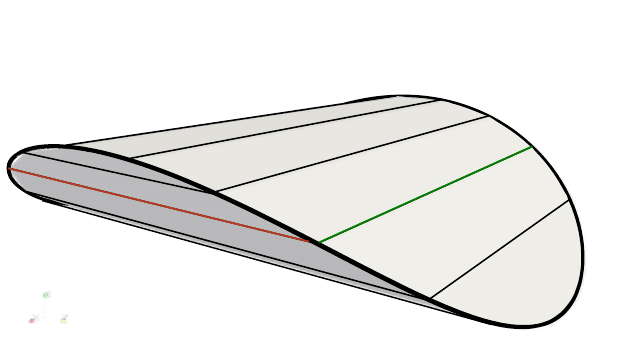}
        \caption{At any given height, the intersection shape has a rectangular cross section. Its width and length are highlighted by the green and red lines.}
        \label{fig:intersectingCylinders}
    \end{figure}
    
    As illustrated in Fig. \ref{fig:intersectingCylinders}, cross-sections of the intersection shape in the $XZ$ plane are rectangles with dimensions $d$ and $w$. 
    These dimensions depend solely on the vertical coordinate $y$.    
    Note that the intersection shape possesses an improper rotational symmetry: reflecting the geometry across the $Y=0$ plane followed by a rotation of $\pi/2$ maps the object onto itself.
    Consequently, the partial volumes and surface areas above and below the $Y=0$ plane are identical.
    
    \begin{figure}[H]
        \centering
        \includegraphics[width=0.9\linewidth]{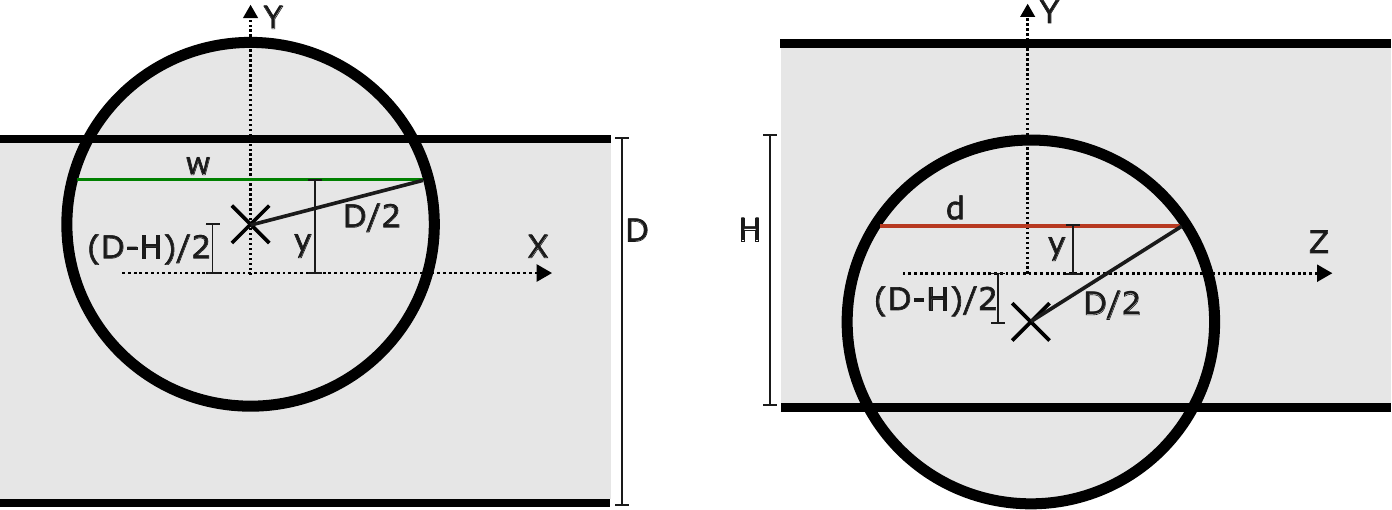}
        \caption{Cross-sectional cuts illustrating the relation of width and height with the intersection volume. The origin of the coordinate system is placed in the center-of-mass of the intersection volume.}
        \label{fig:w_d_sketch}
    \end{figure}

    Figure \ref{fig:w_d_sketch} illustrates how the width and depth depend on $y\in [0,H/2]$. 
    The width and depth are given by:    
    \begin{equation*}
        w(y) = \sqrt{ D^2 - \left( 2y - D + H \right)^2 }
    \end{equation*}
    \begin{equation*}
        d(y) = \sqrt{ D^2 - \left( 2y + D - H \right)^2 }
    \end{equation*}
    Defining the dimensionless width and depth as $w^\prime = w/D$ and $d^\prime = d/D$, respectively, the expressions reduce to:
    \begin{equation} \label{eq:relWidth}
        w^\prime = \sqrt{ 1 - (2y^\prime - 1 + \delta )^2 }
    \end{equation}
    \begin{equation}\label{eq:relLength}
        d^\prime = \sqrt{ 1 - (2y^\prime + 1 - \delta )^2 }
    \end{equation}
    where $y^\prime = y/D$ and $\delta = H/D$.
    The dimensionless volume, normalized by a cube of side length $D$, is obtained by integrating the cross-sectional area $w \cdot d$ along the $y$-axis:

    \begin{equation*}
        \frac{V}{D^3} 
        = \frac{2}{D^3} \int_0^{H/2} dy \, w \, d  
        = 2 \int_0^{\delta/2} dy^\prime \, w^\prime \, d^\prime
    \end{equation*}

    Using equations (\ref{eq:relWidth}), (\ref{eq:relLength}) in the integral above, and $V^\prime = V/D^3$, we arrive at:
    
    \begin{equation} \label{eq:integralSolutionVolume}        
        V^\prime (\delta)
        = 8 \int_0^{\delta/2} dy^\prime \, \sqrt{ 
            y^{\prime 4} - Q_1 y^{\prime 2} + Q_2^2 } ,
    \end{equation}

    with $Q_1 = 1 - \delta + \delta^2/2$ and $Q_2 = \delta/2 - (\delta/2)^2 $. 

    Equation (\ref{eq:integralSolutionVolume}) represents the general solution for the intersection volume. 
    Numerical integration of this expression yields the values listed in Tab. \ref{tab:integralSolutionTable} and the solid curve in Fig. \ref{fig:SurfaceAndVolumeCurves}(left).
    Note that the volume function (\ref{eq:integralSolutionVolume}) is not point symmetric around the half-intersection depth.

    Specifically, in the case of fully intersecting cylinders, the intersection volume integral is readily solved analytically, resulting in $V^\prime (1) = 2/3$.
    The intersection volume of two fully intersecting cylinders of diameter $D$ equals two-thirds of the volume of a cube with side length $D$. 
    This result is also known as the volume of the Steinmetz solid. 

    The intersection volume (\ref{eq:integralSolutionVolume}) is excellently approximated through the empirical relation (\ref{eq:approxSolutionVolume}), as indicated by the dashed curve in Fig. \ref{fig:SurfaceAndVolumeCurves}(left) and the relative errors in Tab. \ref{tab:integralSolutionTable}.

    \begin{equation}\label{eq:approxSolutionVolume}
        V^\prime_{approx}(\delta) = \frac{1}{3} \Big( 1 - \cos( \delta \pi ) \Big)
    \end{equation}

\subsection{Analytical Derivation of Intersection Surface Area}
\begin{figure}[H]
        \centering
        \includegraphics[width=1.\linewidth]{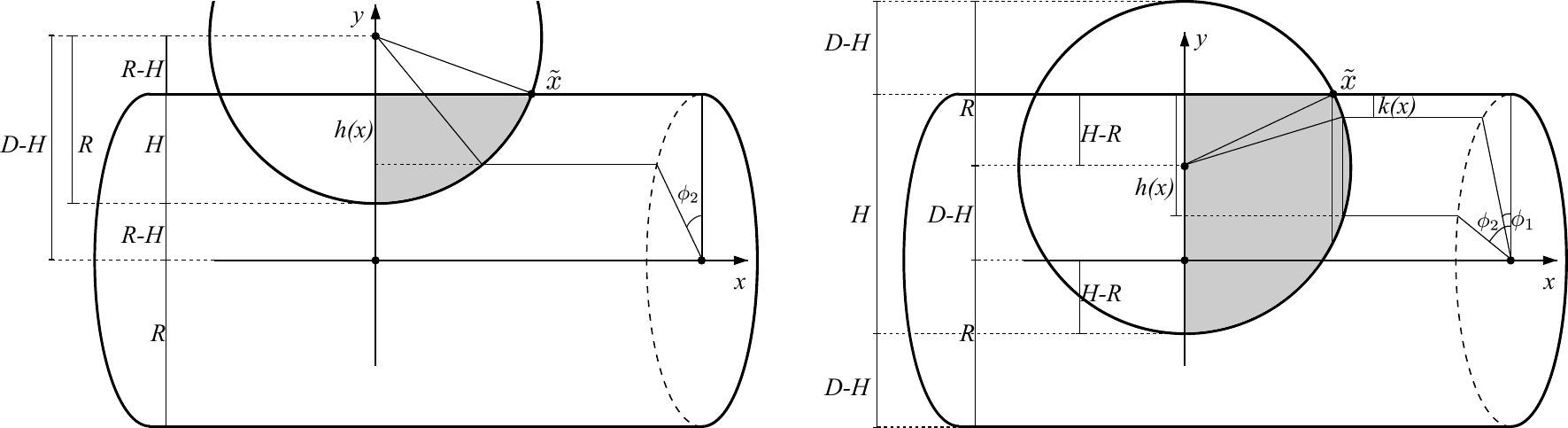}
        \caption{Intersecting cylinders, illustrating geometric features for calculating the gray shell sections. The cylindrical coordinate system is placed in the bottom cylinder. Left: $\delta \leq 1/2$, Right: $\delta > 1/2$}
        \label{fig:areaSketchDrawing}
    \end{figure}

    A cylindrical coordinate system $(r, \phi, x)$, aligned with the bottom cylinder, is defined.
    The surface of the intersection volume consists of two cylindrical shell sections: one belonging to the bottom cylinder and one to the top cylinder, see Fig. \ref{fig:intersectingCylindersSeries}.
    Each cylindrical shell section possesses two planes of symmetry and can be decomposed into four congruent surfaces.
    The gray surface in Fig. \ref{fig:areaSketchDrawing} (facing the reader) is a cylindrical shell section of the bottom cylinder and its surface area is one-eighth of the entire surface area.
    The area of this shell section is calculated as a surface integral in cylindrical coordinates at constant radius $R$:

    \begin{equation*}
        A = 8 R \int_0^{x_{max}}  dr \int_{\phi_1}^{\phi_2} d\phi  
    \end{equation*}
    Along the x-direction, the shell section extends from $0$ to $\Tilde{x}$ for $\delta \leq 1/2$, and from $0$ to $R$ for $\delta > 1/2$.
    At each position $x$ the shell section is spanned between the two azimuth angles $\phi_1, \phi_2$.
    When $\delta\leq 1/2$, $\phi_1=0$, and when $\delta > 1/2$, the value of $\phi_1$ depends on whether $x\leq \Tilde{x}$ or $x>\Tilde{x}$: 
    $\phi_1=0$, if $x\leq \Tilde{x}$ and $\phi_1 \neq 0$, if $x > \Tilde{x}$.

    \begin{align}
        A &= 8 R \int_0^{\Tilde{x}} dr \, \phi_2  && \delta \leq \frac{1}{2} \label{eq:earlySurfaceIntegralsDeltaLEQ05}\\
        A &= 8 R \left( 
        \int_0^{\Tilde{x}} dr \, \phi_2 + 
        \int_{\Tilde{x}}^R dr \, \phi_2 - \phi_1 
        \right) && \delta > \frac{1}{2},\label{eq:earlySurfaceIntegralsDeltaGR05}
    \end{align}
    
    where the position $\Tilde{x}$ is calculated as 
    \begin{equation*}
        \Tilde{x} = \sqrt{2RH-H^2}
        \end{equation*}

    The local intersection depth is denoted by $h(x)$.
    It is related to the position $x$ through $R^2 = x^2 + ( h + R-H)^2$.
    For $\delta \leq 1/2$, $h$ is related to $\phi_2$ through $cos(\phi_2)=(R-h)/R$, and when $\delta > 1/2$, $cos(\pi-\phi_2) = (h-R)/R$.
    In either case, isolating $\phi_2$ yields:
    \begin{equation*}
        \phi_2 = cos^{-1} \left( 1 - \frac{h}{R}   \right) 
        = cos^{-1} \left( 2 - \frac{H}{R} - \sqrt{ 1-\frac{x^2}{R^2}  }   \right)
    \end{equation*}

\begin{figure}[H]
        \centering
        \includegraphics[width=0.49\linewidth]{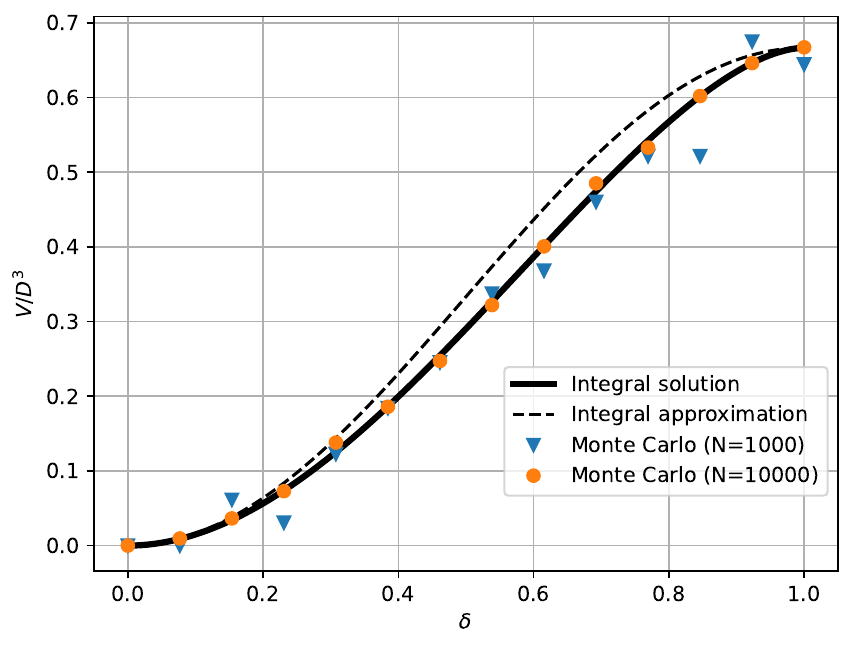}            
        \hfill
        \includegraphics[width=0.49\linewidth]{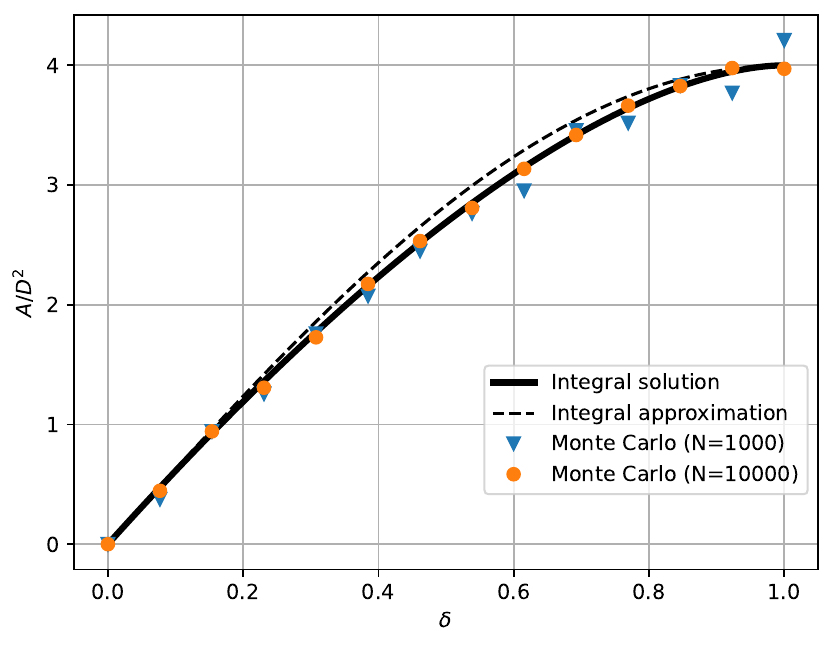}
        \caption{Reduced (volume|surface area) against intersection depth, displaying the integral solution, its empirical approximations, and two Monte Carlo approximations with different number of sample points. Left: reduced volume, Right: reduced surface area.}
        \label{fig:SurfaceAndVolumeCurves}
    \end{figure}

    Similarly, for $\delta > 1/2$ and $x>\Tilde{x}$, the distance $k(x)$ is the local gap between the intersection volume and the top-most position of the bottom cylinder, see Fig. \ref{fig:areaSketchDrawing}, right.
    $k$ and $x$ are related through $R^2 = x^2 + ( H-R-k)^2$, and $k$ and $\phi_1$ are related through $cos(\phi_2)=(R-k)/R$.
    Combining these relations yields:
    \begin{equation*}
        \phi_1 = cos^{-1} \left( 1 - \frac{k}{R}   \right) 
        = cos^{-1} \left( 2 - \frac{H}{R} + \sqrt{ 1-\frac{x^2}{R^2}  }   \right)
    \end{equation*}
    Finally, substituting $H/R = 2\delta$, $x^\prime = x/R$, $A^\prime = A/D^2$, $Q=2-2\delta$, $\Tilde{x}^\prime= \Tilde{x}/R= 2\sqrt{\delta - \delta^2}$ the integrals (\ref{eq:earlySurfaceIntegralsDeltaLEQ05}), (\ref{eq:earlySurfaceIntegralsDeltaGR05}) are written as:
    
    \begin{align}
    A^\prime &=
      2 \int_0^{\Tilde{x}^\prime} dx^\prime cos^{-1} \Big( Q -\sqrt{ 1-{x^\prime}^2 } \Big)    
      &&\delta \leq \frac{1}{2} \label{eq:integralSurfaceSolutionDeltaLEQ05} \\
    A^\prime &=
      2 \left( \int_0^1 dx^\prime cos^{-1} \Big( Q - \sqrt{ 1-{x^\prime}^2 }\Big) - 
    \int_{\Tilde{x}^\prime}^1 dx^\prime cos^{-1} \Big( Q + \sqrt{ 1-{x^\prime}^2 }\Big) \right)
       &&\delta > \frac{1}{2} \label{eq:integralSurfaceSolutionDeltaGR05}
    \end{align}
    The integrals above describe the reduced surface area of the entire intersection shape.
    The integral can be solved analytically only for $\delta = 0$ or $\delta=1$; otherwise, numerical integration is required.
    The resulting reduced surface area is plotted against the intersection depth in Fig. \ref{fig:SurfaceAndVolumeCurves}(right) as a solid curve, and certain calculated integral values are found in Tab. \ref{tab:integralSolutionTable}.

    When $\delta=1$, integral (\ref{eq:integralSurfaceSolutionDeltaGR05}) can be reduced significantly, $Q=0$, $\Tilde{x}^\prime=0$:
    
    \begin{align*}
        A^\prime &=
      2 \left( \int_0^1 dx^\prime cos^{-1} \Big( - \sqrt{ 1-{x^\prime}^2 }\Big) - 
    \int_{0}^1 dx^\prime cos^{-1} \Big( \sqrt{ 1-{x^\prime}^2 }\Big) \right)\\
    &= 2 \left( \pi - 2 \int_{0}^1 dx^\prime cos^{-1} \Big( \sqrt{ 1-{x^\prime}^2 }\Big) \right)\\
    &= 2 \left( \pi - 2 \left[ x^\prime \, cos^{-1} \Big( \sqrt{ 1-{x^\prime}^2 }\Big)    + \sqrt{ 1-{x^\prime}^2} \right]_0^1 \right)\\
    &=4
    \end{align*}
    This solution is the surface area of the Steinmetz solid.

    The surface area integrals (\ref{eq:integralSurfaceSolutionDeltaLEQ05}), (\ref{eq:integralSurfaceSolutionDeltaGR05}) can be excellently approximated with the empirical relation:
    \begin{equation}
        A^\prime_{approx} = 4 \sin\left( \frac{\delta \pi}{2}  \right)
    \end{equation}
    The resulting plot is shown as the dashed curve in Fig. \ref{fig:SurfaceAndVolumeCurves}(right) and the approximation errors are listed in Tab. \ref{tab:integralSolutionTable}.

    \begin{table}[H]
    \centering
    \begin{tabular}{cccccccccc}
    \hline
    $\delta$ & 0.1 & 0.2 & 0.3 & 0.4 & 0.5 & 0.6 & 0.7 & 0.8 & 0.9   \\
    \hline
    $V^\prime$  & 0.015 & 0.056 & 0.120 & 0.199 & 0.290 & 0.387 & 0.481 & 0.567 & 0.635   \\
    $(V^\prime - V^\prime_{approx})/V^\prime  (\%) $  & -9.4 & -12.8 & -14.8 & -15.4 & -14.8 & -12.9 & -9.9 & -6.3 & -2.4 \\
    $A^\prime$ & 0.612 & 1.190 & 1.732 & 2.232 & 2.688 & 3.093 & 3.440 & 3.719 & 3.916  \\
    $(A^\prime - A^\prime_{approx})/A^\prime (\%) $  
    & -2.2 & -3.8 & -4.9 & -5.3 & -5.2 & -4.6 & -3.6 & -2.3 & -0.9  \\
     \hline
    \end{tabular}
    \caption{Integral values of reduced volume and surface area, together with relative errors of the respective approximations.}\label{tab:integralSolutionTable}
    \end{table}

\section{Quasi-Monte Carlo Estimation}
    In this section, a Quasi-Monte Carlo method to estimate the intersection volume and surface area of two arbitrary finite cylinders is described. 
    The corresponding Python implementation is provided in the Appendix.
    All geometric parameters (endpoints of axes, radii) can be freely specified, allowing application to perpendicular, skew, or partially overlapping cylinders.

    The core of the algorithm involves calculating the shortest Euclidean distance from sample points $P_{i}$ to a cylinder axis defined by endpoints $A$ and $B$:
    
    \begin{equation*}
        \overrightarrow{AP}_{i} = P_{i} - A    
    \end{equation*}

    \begin{equation*}
        \overrightarrow{AB} = B - A    
    \end{equation*}
    
    \begin{equation*}
        t_{i} = \frac{\overrightarrow{AP}_{i} \cdot \overrightarrow{AB}}{|\overrightarrow{AB}|^{2}}        
    \end{equation*}

    \begin{equation*}
        t_{i}^{'} = \text{clip}(t_{i},\ 0,\ 1)    
    \end{equation*}
    
    \begin{equation*}
        Q_{i} = A + t_{i}^{'}\overrightarrow{AB}    
    \end{equation*}

    \begin{equation*}
        d_{i} = \| P_{i} - Q_{i}\|    
    \end{equation*}

    Here, $Q_{i}$ is the point on the axis segment closest to $P_{i}$, while $d_{i}$ is the orthogonal distance. 
    The scalar $t_{i}^{'}$ represents the relative position of the closest point on the axis segment $AB$ to the point $P_{i}$, expressed as a fraction between 0 and 1 along the length of the segment. 
    This is an established geometric method (\cite{ericson2005}).
    
    Samples are generated using a Sobol sequence, a low-discrepancy quasi-random generator.

\subsection{Volume}
    
    For volume estimation, points are drawn in the interior using cylindrical coordinates, transformed back to Cartesian space via an orthogonal local basis $(\mathbf{u},\mathbf{v},\mathbf{d})$, with $\mathbf{d}$ along the cylinder axis and $\mathbf{u},\mathbf{v}$ constructed orthogonal to $\mathbf{d}$ via cross products. Radial samples are drawn by taking the square root of uniformly distributed random numbers, which ensures that points are uniformly distributed across the circular cross-section.
    
    The intersection volume estimate proceeds as follows:
    
    \begin{enumerate}
        \item Generate $n$ samples within the first cylinder's interior.
        \item For each sample, compute its orthogonal distance to the second cylinder's axis.
        \item Mark the sample as ``hit'' if this distance is less than or equal to the second cylinder's radius.
        \item The estimated intersection volume is
        \[
        V_{\text{est}} = \left( \frac{\text{hits}}{n} \right) V_{\text{cyl 1}}
        \]
        where $V_{\text{cyl 1}} = \pi r_{1}^{2}\ L_{1}$.
    \end{enumerate}
    
\subsection{Surface area}

    For surface area estimation, samples are instead mapped to the lateral surface using the same basis, with the angular and longitudinal coordinates spanning the full $2\pi$ and axis length, respectively.
    
    For surface area estimation:
    
    \begin{enumerate}
        \item Generate $n$ samples on the surface of each cylinder individually.
        \item For each sample, compute its distance from the axis of the other cylinder.
        \item For each cylinder, the proportion of its surface points found to lie inside the other cylinder is used to estimate the portion of the lateral surface that forms part of the intersection.
        \item The estimated intersection surface area is:
        \[
        A_{\text{est}} = f_{1}A_{\text{cyl 1}} + f_{2}A_{\text{cyl 2}}
        \]
        where $f_{1,2}$ are the fractions of surface samples on each cylinder found to lie inside the other, and $A_{\text{cyl}} = 2\pi rL$.
    \end{enumerate}

\bibliography{references}

\newpage
\section*{Appendix: Monte Carlo implementation}
\begin{lstlisting}
import numpy as np
from scipy.stats import qmc


def point_to_segment_distance_vectorized(P, A, B):
    AP = P - A
    AB = B - A
    AB_dot = np.dot(AB, AB)
    t = np.einsum('ij,j->i', AP, AB) / AB_dot
    t = np.clip(t, 0, 1)
    closest = A + t[:, np.newaxis] * AB
    return np.linalg.norm(P - closest, axis=1)


def cylinder_surface_area(A, B, r):
    return 2 * np.pi * r * np.linalg.norm(B - A)


def sample_cylinder_surface(A, B, r, n_samples):
    axis = B - A
    length = np.linalg.norm(axis)
    direction = axis / length if length > 0 else np.array([0, 0, 1])

    # Create orthogonal basis
    if np.allclose(direction, [0, 0, 1]):
        u = np.array([1, 0, 0])
    else:
        u = np.cross(direction, [0, 0, 1])
        u /= np.linalg.norm(u)
    v = np.cross(direction, u)

    # Generate Sobol samples
    m = int(np.ceil(np.log2(n_samples)))
    sampler = qmc.Sobol(d=2, scramble=True)
    samples = sampler.random_base2(m=m)[:n_samples]

    # Convert samples to cylindrical coordinates
    theta = samples[:, 0] * 2 * np.pi
    z = samples[:, 1] * length

    # Convert to 3D coordinates with vectorized operations
    points = (
        A[:, None] +
        z[None, :] * direction[:, None] +
        r * np.cos(theta)[None, :] * u[:, None] +
        r * np.sin(theta)[None, :] * v[:, None]
    )
    return points.T


def sample_intersection_volume(A1, B1, r1, A2, B2, r2, n_samples=5_000_000):
    axis = B1 - A1
    length = np.linalg.norm(axis)
    direction = axis / length if length > 0 else np.array([0, 0, 1])

    # Create orthogonal basis vectors for Cylinder 1's coordinate system
    if np.allclose(direction, [0, 0, 1]):
        u = np.array([1, 0, 0])
        v = np.array([0, 1, 0])
    else:
        u = np.cross(direction, [0, 0, 1])
        u /= np.linalg.norm(u)
        v = np.cross(direction, u)

    # Generate samples
    sampler = qmc.Sobol(d=3, scramble=True)
    samples = sampler.random_base2(m=int(np.log2(n_samples)))

    # Convert to cylindrical coordinates with uniform volume distribution
    radial = r1 * np.sqrt(samples[:, 0])  # Correct sqrt for uniform density
    theta = 2 * np.pi * samples[:, 1]
    z_local = length * samples[:, 2]

    # Convert to 3D points in world coordinates
    points = (
            A1[:, None] +
            z_local[None, :] * direction[:, None] +
            radial[None, :] * (np.cos(theta)[None, :] * u[:, None] +
                               np.sin(theta)[None, :] * v[:, None])).T
    d2 = point_to_segment_distance_vectorized(points, A2, B2)
    hit_ratio = np.mean(d2 <= r2)
    cyl_volume = np.pi * r1 ** 2 * length

    return hit_ratio * cyl_volume


def sample_intersection_surface_area(A1, B1, r1, A2, B2, r2, n_samples=5_000_000):
    # Sample both surfaces
    points1 = sample_cylinder_surface(A1, B1, r1, n_samples)
    points2 = sample_cylinder_surface(A2, B2, r2, n_samples)

    # Check containment using vectorized distances
    in_cyl2 = point_to_segment_distance_vectorized(points1, A2, B2) <= r2
    in_cyl1 = point_to_segment_distance_vectorized(points2, A1, B1) <= r1

    # Calculate estimated areas
    area1 = cylinder_surface_area(A1, B1, r1)
    area2 = cylinder_surface_area(A2, B2, r2)

    return (np.mean(in_cyl2) * area1) + (np.mean(in_cyl1) * area2)

\end{lstlisting}

\end{document}